\documentclass[preprint,showpacs,amsmath,amssymb]{revtex4} 
\usepackage{graphicx}
\usepackage{psfrag}
\begin{document}
%%%%%%%%%%%%%%%%%%%%%%%%%%%%%%%%%%%%%%%
%%%%%%%%%%%%%%%%%%%%%%%%%%%%%%%%%%%%%%%
\title{Giant diffusion and coherent transport in tilted periodic 
 inhomogeneous systems.}
\author{Debasis Dan}
\email{dan@iopb.res.in}
\author{A. M. Jayannavar }
\email{jayan@iopb.res.in}
\affiliation{Institute of Physics, Sachivalaya Marg,
           Bhubaneswar 751005, India.}
\date{\today}
%%%%%%%%%%%%%%%%%%%%%%%%%%%%%%%%%%%%%%%%
%%%%%%%%%%%%%%%%%%%%%%%%%%%%%%%%%%%%%%%%
\begin{abstract}
  We investigate the dynamics of an overdamped Brownian
particle moving in a washboard potential with space dependent friction 
coefficient. Analytical expressions have been obtained for current and 
diffusion coefficient. We show that 
the effective diffusion coefficient can be enhanced or suppressed  
compared to that of
the uniform friction case. The diffusion coefficient is maximum near the
critical threshold ($F_{c}$), which is sensitive to temperature and
the frictional profile. In some parameter regime we observe that increase 
in noise (temperature) decreases the diffusion, which is counter-intuitive. 
This leads to coherent transport with large mean velocity accompanied by small
diffusion. This is shown explicitly by analysis of P\'{e}clet number, which 
has been introduced to study coherent or optimal transport. This
phenomena is complementary to giant diffusion.    
\end{abstract}
\maketitle

\section{Introduction}

       In recent times there has been a renewed interest in the study
of transport properties of Brownian particles moving in periodic
potential \cite{Ris}, with special emphasis on coherent transport and giant
diffusion \cite{Main,March}. 
%%Two effects namely giant diffusion and coherent transport%%  
%%have attracted attention in recent years.%%% 
The phenomenon of coherent or
optimal transport is complimentary to the enhanced diffusion, wherein
one is mainly concerned with transport currents with minimal
dispersion or diffusion \cite{Fnl}. Compared to free diffusion 
coefficient(DC, $D = k_{B}T/ \gamma$), DC  is  suppressed 
in the 
presence of periodic potential. However, in a nonequilibrium case 
i.e., in the presence of bias, it has been recently shown that DC can be made
arbitrarily large (giant diffusion) compared to the bare diffusion,  in 
the presence of periodic potential \cite{Main}. This enhancement at low 
temperature takes place near the critical threshold ( at which 
deterministically running solution sets in). The reason 
for this enhancement has been attributed to the existence of instability 
between locked to running solution.  In some cases enhancement by
fourteen order of magnitude has been predicted, so that diffusion can 
be observed on a macroscopic scale at room temperature
\cite{Main}. This enhancement  
decreases as we move away from the critical field in either direction. 
Exact result for DC in arbitrary potentials has been obtained in term 
of quadratures. In special cases an elegant simplification of 
quadrature have been carried out. Near the critical tilt, scaling behavior 
of DC for weak thermal noise has been obtained and different universality
classes have been identified \cite{Main}. Approximate expression for DC
in terms of mobility 
has been obtained earlier which deviates from the exact results near
the critical threshold \cite{March}.   

     In a related development, study of coherent or optimal transport 
has been reported recently \cite{Fnl}. Coherent or optimal transport of 
Brownian particles refer to the case
of large mean velocity accompanied with minimal diffusion. This can be 
quantified by dimensionless P\'{e}clet number (ratio of mean velocity to DC). 
The transport is most coherent when this number is maximum. The particle
motion is mainly determined by two time scales; noise driven escape from
potential minima over the barrier along the bias, followed by the relaxation
into the next minima. The former depends strongly on temperature and 
the later weakly on the noise strength and has a small variance. It is 
possible to obtain coherent transport in the parameter
regime at which the traversal time across the two consecutive minima
in a washboard potential is dominated by the relaxational
time.  At optimal
noise intensity certain regularity of the particle motion is expected
which accounts for the maximum of P\'{e}clet number. In some
cases (molecular separation devices) for higher reliability, one
requires higher currents but with less dispersion (or DC)
~\cite{Geier}.  This effect of 
coherent transport is related to the phenomenon of coherence resonance
~\cite{Cr} observed in  excitable systems~\cite{Fnl}.

     In the present work we study both the mentioned phenomena in a
space dependent frictional medium. For this we have considered a
simple minimal model where the potential is sinusoidal and 
the friction 
coefficient is also periodic (sinusoidal) with the same period, but
shifted in phase. Frictional inhomogeneities are not
uncommon in nature. Here we mention a few.   
Brownian motion in confined geometries, porous media experience space dependent
friction~\cite{Conf}. 
Particles diffusing close to surface have space dependent
friction coefficient~\cite{Conf,Surf}. It is believed that molecular motor
proteins move
close along the periodic structure of microtubules and will therefore
experience a position dependent friction ~\cite{Spc_mob}. 
Inhomogeneities in mobility ( or friction) occurs naturally in super
lattice structures and Josephson 
junctions~\cite{Falco}.  

                                    Frictional inhomogeneity
changes the dynamics of the diffusing particle non-trivially, thereby
affecting the passage times in 
different regions of the potential. However, it does not effect the
equilibrium distribution. Thus thermally activated
escape rates and relaxational rates within a given spatial period are affected
significantly.  This in turn has been shown to give rise to  several
counter-intuitive 
phenomena. Some of them are stochastic resonance in absence of
external periodic drive \cite{Sr}, noise induced stability in 
washboard potential \cite{Dan}. Single and multiple current reversals in
adiabatic ~\cite{Danphys}  and nonadiabatic  rocked system (thermal ratchets ~\cite{Ratchet})
respectively have also  been reported \cite{Dan2}. In  
these ratchet systems the magnitude of efficiency of energy
transduction in finite
frequency regime may be more than the efficiency in the adiabatic
regime, i.e, quasistatic operation may not be
efficient for conversion of input energy into mechanical work \cite{Dan3}.  All
these above features are absent in the corresponding homogeneous case
for the same simple potential. 

                             In our present work we show that
frictional inhomogeneities can give rise to additional new features in a
tilted periodic potential. The observed giant enhancement of DC near
the critical tilt can be controlled (enhanced or suppressed) in a
space dependent frictional medium by
suitably choosing the phase difference between the potential and the
frictional profile. The most surprising feature is the noise
induced suppression of diffusion,  leading to coherent transport. Our
results are based on 
analytical expressions for P\'{e}clet number and DC in term 
of moments of first passage times. 

         In Section \ref{model} we present our model and derive the
expression for moments of first passage times. Using these, DC and
P\'{e}clet number have been defined.  In section
\ref{DC} we analyze the nature of DC  as function of system
parameters, such as the applied external force and temperature. Section
\ref{peclet} is devoted to the study of coherent or optimal transport in
different regimes of parameter space. Finally in
section \ref{conclusion} we present the summary of our findings. 

\section{Model \label{model}}

            We consider an overdamped Brownian particle moving in a
symmetric one dimensional periodic potential $V_{0}(x)$ with space
dependent friction 
coefficient $\eta(x)$ under the influence of constant external tilt
$F$ at temperature $T$. For simplicity we take $V_{0}(x) = -\sin(x)
\mbox{ and } \eta(x) = \eta_{0}(1 - \lambda \sin(x+2\pi \phi)), \mbox{ where }
|\lambda| < 1$.  $\phi$ determines the relative
phase shift between  friction coefficient and potential. The correct Langevin
equation for such systems has been derived earlier from microscopic
considerations \cite{Langvn} and is given by
%%%%%%%%%%%% 
%
\begin{equation}
 \dot{x} =  -\frac{(V'_{0}(x) - F)}{\eta (x)} - k_{B}T \frac{\eta '(x)}{(\eta
   (x))^{2}} + \sqrt{\frac{k_{B}T}{\eta (x)}}\xi (t) ,
\label{Langvn}
\end{equation}
%
%%%%%%%%%%%%%%%
where $\xi(t)$ is a zero mean Gaussian white noise with correlation
$\left<\xi (t) \xi (t') \right> = 2 \delta(t-t')$. It should be
noted that the above equation involves a multiplicative noise with an
additional temperature dependent drift term which turns out to be
essential for the system to approach correct thermal equilibrium
state in absence of nonequilibrium forces \cite{Sancho}. The quantity
of central 
interest in this work is the effective diffusion coefficient $D$ given 
by \\
\begin{equation}
  D = \lim_{t -> \infty} \frac{\left< x^{2}(t) \right> - \left< x(t)
    \right>^{2}}{2t}. 
\end{equation}
In the absence of potential, $D = \frac{k_{B}T}{\eta}$ (uniform
$\eta$), is the usual Einstein relation. An exact analytical expression 
for $D$ and average current $J$ in terms of the moments of first passage
time have been recently given \cite{Main,Fnl}. If the n-th moment of
the first passage time from an arbitrary 
point $x_{0} \mbox{ to } b$ is given by $T_{n}(x_{0}
\rightarrow b) =  \left<t^{n}(x_{0} \rightarrow b)\right>$, then  
\begin{equation}
     D = \frac{L^{2}}{2} \frac{T_{2}(x_{0} \rightarrow x_{0}+L) -
       [T_{1}(x_{0} \rightarrow x_{0}+L)]^{2}}{[T_{1}(x_{0}
       \rightarrow x_{0}+L)]^{3}} 
\label{diffu}
\end{equation}
                For our problem (\ref{Langvn}), the moments of
first passage time follow closed recursion relation ~\cite{Main,Gard}. 
\
\begin{equation}
  \label{moment}
  T_{n}(x_{0} \rightarrow b) = \frac{n}{D_{0}}\int^{b}_{x_{0}} dx
  \hat{\eta}(x) e^{\frac{{V(x)}}{k_{B}T}}\int^{x}_{\infty} dy
  e^{\frac{-V(y)}{k_{B}T}} T_{n-1}(y \rightarrow b),   
\end{equation}
with $T_{0}(y \rightarrow b) =1$. Here $V(x) = V_{0}(x) - Fx, D_{0} =
k_{B}T/\eta_{0} \mbox{ and } \hat{\eta} = \eta(x)/ \eta_{0}$.  $\eta_{0}$
also happens to be the average value of friction coefficient over a period.  

By using Eq. (\ref{diffu}) and
Eq. (\ref{moment}), and some straight forward algebra, we get
\begin{equation}
  \label{quad}
  D = D_{0}\frac{\int_{x_{0}}^{x_{0}+L} dx
    I_{\pm}(x)I_{+}(x)I_{-}(x)}{[\int_{x_{0}}^{x_{0}+L} dx I_{\pm}(x)]^{3}},
\end{equation}
where, 
%%%%%%
%%% CHECK THE LIMITS
%%%%%%%%%
\begin{subequations}
 \label{I+-}
 \begin{equation}
   I_{+}(x) = \frac{1}{D_{0}} \hat{\eta}(x)e^{\frac{V(x)}{k_{B}T}}
   \int_{x-L}^{x} dy e^{\frac{-V(y)}{k_{B}T}}, 
 \end{equation}
 \begin{equation}
   I_{-}(x) = \frac{1}{D_{0}} e^{\frac{-V(x)}{k_{B}T}}
   \int_{x}^{x+L} dy \hat{\eta}(y) e^{\frac{V(y)}{k_{B}T}}.
 \end{equation}
\end{subequations}
%%%%%%%%%%%%%%%%%%%%%%%%%%%%%%%%%%%%
The average current density $J$ for this system has been derived
earlier~\cite{Dan}
which in term of Eqs. (\ref{I+-}) is given by
\begin{equation}
  \label{curr}
  J = L\frac{1-\exp(-LF/kT)}{\int^{x_{0}+L}_{x_{0}} \frac{dx}{L}I_{\pm}(x)}.
\end{equation}
The above expressions go over to the results obtained earlier for the
case of space independent friction ($\eta(x) = \eta_{0}, \lambda =
0$) \cite{Main}. It should be noted that 
Eqs. (\ref{I+-}) are applicable when $\eta(x) \mbox{ and } V(x)$ have
the same periodicity. Otherwise they have to be modified appropriately. We
obtain results for DC by numerically integrating
Eqs. (\ref{I+-}) using a globally adaptive scheme based on
Gauss-Kronrod rules.
For the special case of $V_{0}(x) = 0$ we get
\begin{equation}
   D = D_{0}(1+\frac{\lambda^{2}}{2}\frac{1}{[(\frac{k_{B}T}{F})^{2}+1]}).
\end{equation} 
We would like to mention here that in the absence of potential, DC
explicitly depends on system inhomogeneities ( via $\lambda$),
however, steady current is independent of $\lambda$ for the same case
~\cite{Dan}. 
$F = 0$ is the equilibrium situation and as expected $D = D_{0}$,
which corroborates with the fact that frictional inhomogeneities 
cannot affect the equilibrium properties of the system. In the
high temperature regime, $D = D_{0}$ as anticipated. For asymptotically 
large field, DC saturates to a $\lambda$ dependent value. This
is solely attributed to space dependent friction. This somewhat
surprising result also appears  in the dependence of current on
$\lambda$ in the presence of potential and high field limit
~\cite{Danphys,Dan2}. 

   In our subsequent analysis all the physical quantities are
expressed in term of 
dimensionless units, DC is scaled with respect to $D_{0}$ or
$V_{0}/\eta_{0}$ and $T$ is
scaled with respect to $V_{0}$, where $V_{0}$ is half the potential
barrier height (which is
one). Period of the potential, $L = 2\pi$.    
  
\section{Results and Discussions \label{results}}
\subsection{Diffusion Coefficient \label{DC}}
         Though the system response to the applied field is generally
given by the stationary current density $J = L<v>$, but this
directed motion (or average position of the particle) is accompanied
by dispersion due to the inherent stochastic 
nature of the transport. It has been shown previously that in a
homogeneous medium this dispersion or diffusion becomes
very large (giant enhancement of DC) near the critical tilt. This
enhancement in DC can be order of magnitude larger than the bare DC
in the absence of potential. We make a
systematic study of this phenomena in the presence of
system inhomogeneities. Though our parameter space is large, we 
restrict to a narrow relevant domain where we observe effects
which are surprising, and arise due to system
inhomogeneities.    

\begin{figure}[h]
\includegraphics[width=8.6cm]{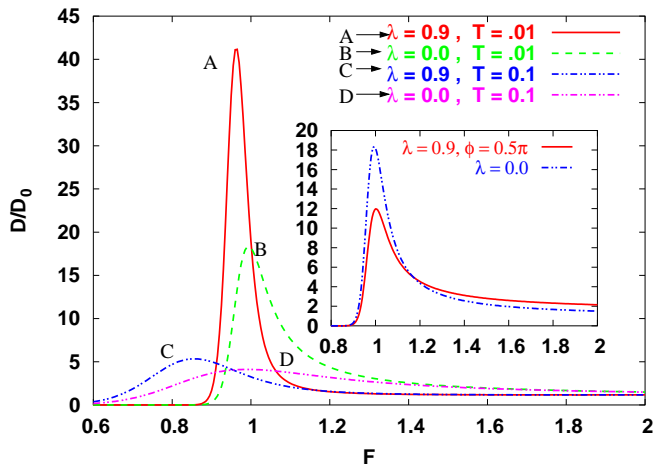}
\caption{\label{F-D}Diffusion coefficient (5) \textit{vs} $F$ for 
  $\phi = 1.6\pi$. Inset shows the suppression of DC for inhomogeneous
  systems($\lambda=0.9, \phi=0.5\pi$, lower curve) as compared to the homogeneous systems.}
\end{figure}
        In fig.~\ref{F-D} we plot DC 
as function of external tilt $F$ for different values of temperature
$T$ ($\lambda = 0.9 \mbox{ and } \phi = 1.6\pi$). It can be seen from 
the figure that DC exhibits a maxima as function of $F$. However,
quantitative details depend sensitively on system parameters such as
$\phi, T \mbox{ and } \lambda$. It can
readily noticed from the curves A and B that DC has been enhanced by
more than factor $2$ as compared with the homogeneous case. On lowering
the temperature the relative enhancement of DC still increases. DC can
also be suppressed by properly tuning $\phi$.  For $\phi =
0.5\pi$, DC near the critical field is suppressed by almost a factor
$2$ as compared to the homogeneous case (
shown in the inset). Thus one can enhance or suppress DC near critical
field by appropriately choosing the system parameters. It should be noted that 
in the case of enhancement, friction coefficient is lower on the
immediate left side of the barrier and higher on the opposite
side, where the relaxational motion takes place. When the frictional profile
is opposite to the above case, suppression in DC
occurs. When the phase difference between the two frictional profiles
differ by $\pi$, then in one case enhancement and in the other 
suppression of DC can be observed as compared with the homogeneous
case.  Thus $\phi$ controls the fluctuations of first passage times,
hence DC and current. 

                       Unlike the behavior in the homogeneous case  
($\lambda = 0$) where for small values of $T$ this peak value occur
exactly at $F = 1$ (critical field), here the peak position is very
sensitive to $\phi$ and can be shifted to either side of the bare
critical field.  Higher the temperature larger is the 
deviation from the critical threshold. 
\begin{figure}[h]
\includegraphics[width=8.8cm]{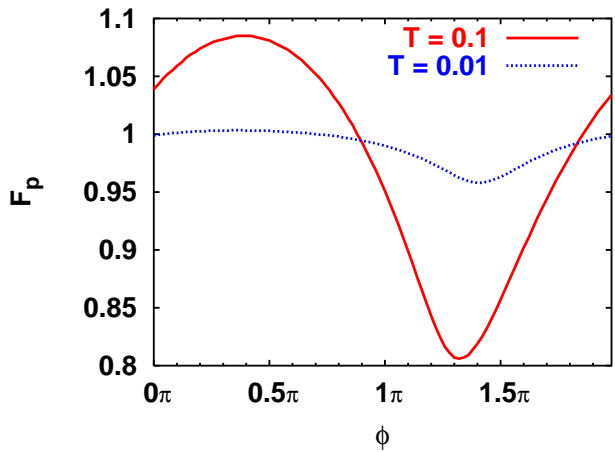}
\caption{\label{thrshld}The value of $F$ at which $D/D_{0}$ peaks ($F_{p}$) versus
  $\phi$ at $\lambda = 0.9$. }
\end{figure}
Fig. \ref{thrshld} shows this
behavior, where we have plotted the force at which diffusion peaks
($F_{p}$) as function of $\phi$ at $\lambda = 0.9 \mbox{ for }T = 0.1 \mbox{ and } 0.01$. For $T = 0.1$ the peak can occur at as low as
$F = 0.8$. The fact that $F=0.8$ is away from critical threshold, 
hence enhancement in DC in this regime has to be attributed to 
space dependent friction. This is a clear example of system
inhomogeneity affecting the dynamical evolution of the particle
nontrivially.  This will be discussed at the end of this section to
explain many of our observations. We would also like to emphasize  
that critical threshold is not altered at temperature $T=0$ for our
present case as shown earlier ~\cite{Dan}. 

\begin{figure}[h]
\includegraphics[width=8.6cm]{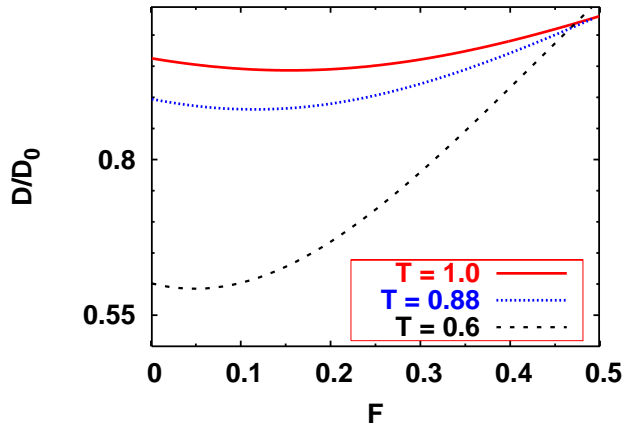}
\caption{\label{min_F}$D/D_{0}$ \textit{vs} $F$ for $\lambda = 0.9 \mbox{ and } \phi = 0.84\pi$. The 
 figure highlights the minima in DC with $F$. } 
\end{figure}

    Since with increasing tilt the barrier to forward motion decreases 
(thereby reducing the effect of exponential suppression of DC in the
presence of periodic potential), therefore it is natural to  
expect that $D/D_{0}$ will increase with increasing $F$ (for $F <
\mbox{ barrier height }$). This is amply reflected in ref. \cite{Main},
which corresponds to  
our $\lambda = 0$ case. In the presence of space dependent friction
($\lambda \neq 0$), $D/D_{0}$ can show a minimum (as shown in
fig. ~\ref{min_F}) with increasing force 
( for $F < \mbox{ critical field }$). This is  surprising given the
fact that current increases monotonically with increasing field (which
we have checked separately) as expected. To observe this phenomena one 
has to properly choose the parameters. This is akin to the phenomena 
of coherent transport, wherein, increase in current is
accompanied by decrease in DC. This we have discussed in detail in the
later section as function of $T$, however, it is observed here as
function of $F$ also.

\begin{figure}[h]
\includegraphics[width=8.4cm]{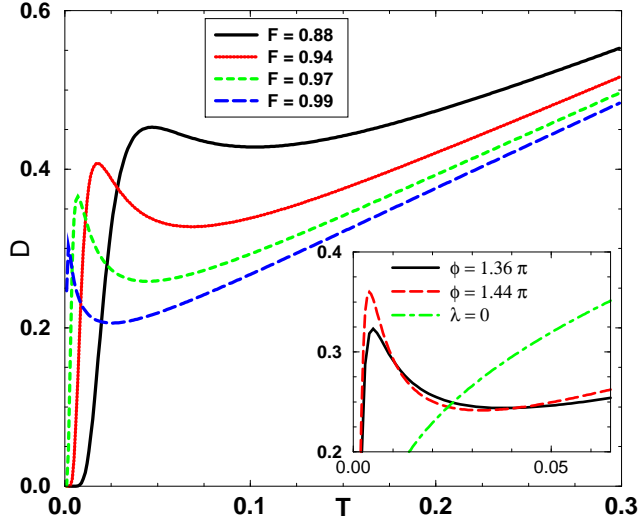}
\caption{\label{minima_T} D \textit{vs} $k_{B}T$ for $\lambda = 0.9
  \mbox{ and }\phi = 1.4\pi$ and 
  various values of F. The inset compares the variation of D with $k_{B}T$ for
  $\lambda = 0.9$ with $\lambda = 0.0$. $F = 0.98$ in the inset.}
\end{figure}
     Next we proceed to present the most interesting consequence of
space dependent friction coefficient in our simple model. Unlike the expected
result where the diffusion coefficient increases with 
temperature, here the diffusion can be suppressed by
increasing the temperature. Fig. \ref{minima_T} highlights this. Here
we have plotted $D$ 
as function of $T$. $D$ is scaled with $V_{0}/\eta_{0}$ as
mentioned in the Sec. \ref{model}. For the inset we have taken 
$\phi = 1.36 \pi \mbox{ and } 1.44 \pi 
\mbox{ at } F = 0.98 \mbox{ and } \lambda = 0.9$.  For homogeneous
case ($\lambda = 0$), 
the minima is absent and $D$ increases monotonically with
$T$. However, minima is clearly seen for the case $\lambda \neq
0$. The observed suppression occurs when $F$ is close to the 
critical field. In can be clearly seen from
the figure that as we go away from the critical field the minima in
DC shifts to higher values of temperature and importantly it
becomes shallower. Below certain value of $F$, minima and hence the
suppression of DC disappears. The existence of the suppression in DC
is attributed to the competition between two time scales. First, noise
driven escape over potential barrier from the minima along the bias
and the second time scale being the relaxation into the next potential 
well from the barrier top ~\cite{Fnl}. It has been argued before that
the second time scale is weakly dependent on noise strength and has a
small variance as opposed to the first one. It is obvious that in
transport processes when
the second time scale dominates over the first it is expected to
result in suppression of DC as function of noise strength (see for
details ref. \cite{Fnl}). In the high temperature regime large thermal 
noise leads to large variance in the second time scale as the random motion
of the particle both along and against the bias becomes equally
important. Thus the DC 
increases as expected for higher temperatures, hence minima in
DC. It is to be noted that for the case where $\phi = 1.4\pi$, the
friction coefficient $\eta(x)$
is smaller near the barrier heights and moreover $F$ being close to
critical threshold makes Arrhenius barrier crossing time scale (first 
time scale) smaller. As opposed to this, $\eta(x)$ is higher between
the barrier height and the next potential minima along the bias thus
slowing down the relaxation motion to the next minima. This naturally
enhances the dominance of the relaxation time scale over the barrier
crossing rate. This qualitatively explains our observed
behaviour. 

\begin{figure*}[h]
\includegraphics[width=10.6cm]{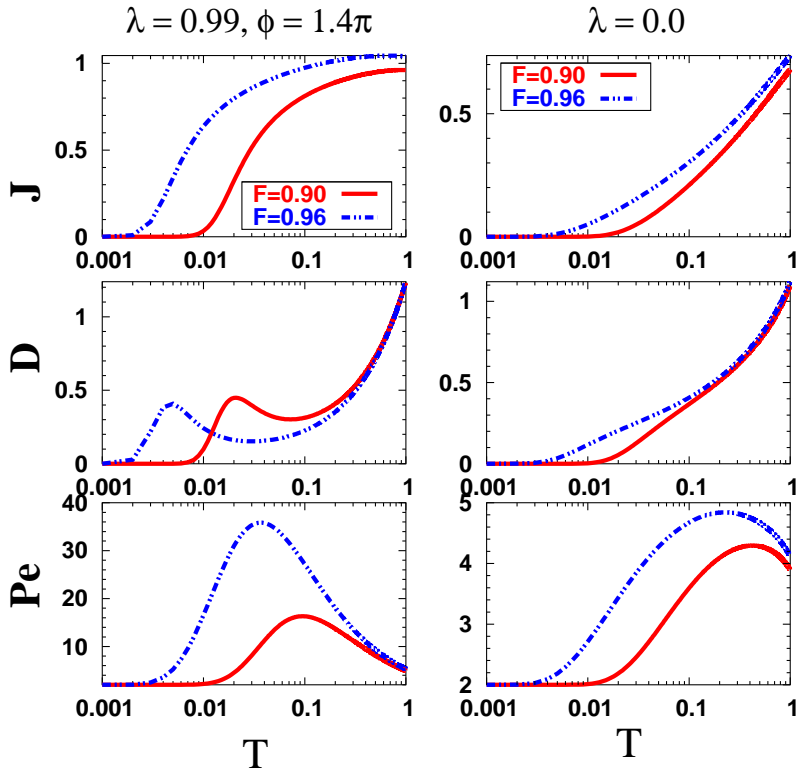}
\caption{\label{peclet_color} The figure compares current, DC and P\'{e}clet number ( from top
 to bottom) of the inhomogeneous system ($\lambda = 0.99 \mbox{ and }\phi = 1.4
\pi$, left hand side figures) with that of the homogeneous 
 system ($\lambda = 0$, right hand side figures).}
\end{figure*}
%%%%%%%%%%%%%%%%%%%%%%%%%%%%%%%%%%%%%%%%%%%%%
\subsection{\label{peclet} P\'{e}clet number and coherent motion}

    By coherent motion we mean large particle current with minimal
diffusion. This property can be quantified by 
the dimensionless P\'{e}clet number $Pe$ defined as ~\cite{Fnl} 
\begin{equation}
Pe = \frac{L \left< \dot x \right>}{D},
\end{equation}
where $L \left< \dot x \right>$ is the average current density $J$. The
expression for the current density is given in Eq. (\ref{curr}). We make
use of  expressions (\ref{quad} and \ref{curr}) to calculate $Pe$. 
The parameter values at which $Pe$ shows maxima correspond to the
most coherent motion for that particular model. Higher the $Pe$ more
coherent is the 
transport. It should be noted that the $Pe$ can show maxima though
neither $J$ nor $D$ may show extrema. In fig.~\ref{peclet_color}, we
plot the current $J$, diffusion coefficient $D$ and the P\'{e}clet number
$Pe$ (from top to 
bottom) for both space dependent (left column of fig.~\ref{peclet_color})and space
independent friction (right column of fig.~\ref{peclet_color})
cases. The average friction coefficient over a period in inhomogeneous 
case equals to the value taken for the homogeneous case. As pointed
out earlier, though in the homogeneous case 
 $J \mbox{ and } D$ is monotonic increasing function of $T$, the
p\'{e}clet number shows a maxima, the maximum value being very small
compared to the space dependent friction case. For the chosen
parameter values in the space dependent friction case,  DC shows a
minima with $T$ while the current increases monotonically. Moreover
the magnitude of this current is larger than the corresponding current
in the homogeneous friction case. This is 
indeed the most favorable condition of transport where increasing
current is accompanied by decreasing DC. This is aptly reflected in
the $Pe$, which shows enhancement (coherent motion or
optimal transport) by an order of
magnitude as compared with corresponding uniform friction case. The
P\'{e}clet number is 
sensitively dependent on the phase factor $\phi$ and it can also be
suppressed (diffusion dominates the transport) which we have not
reported here. Hence we can
control the degree of coherent motion. 
\section{\label{conclusion} Conclusions}
%%%              
                 We have thus shown that both giant diffusion and
coherent transport in a tilted periodic potential is sensitive to the
frictional inhomogeneities of the medium. To analyze this problem we
have taken a simple sinusoidal potential and periodic frictional
profile with same periodicity but with a phase difference. Depending
on the system parameters the value of DC near the critical threshold
and P\'{e}clet number (indicating coherence in the transport) can be
enhanced or decreased by an order of magnitude. Both these
complimentary effects are important for practical applications
~\cite{Main,Geier}. The regime where we 
observe the optimal transport is accompanied by decrease of DC with
temperature, the aspect which is absent in the
corresponding homogeneous case. We have focussed on a restricted
parameter regime to highlight the most interesting results arising 
due system inhomogeneities, in systems with simple potential. It is
known that in the present model depending on system parameters current
decreases with temperature, the effect akin to noise induced stability
~\cite{Nis}. However, in this regime we have not observed any dramatic
effect on DC as well as P\'{e}clet number. It is not clear whether noise
induced stability (NIS) can enhance the coherence in the motion. The
phenomenon of stochastic resonance (SR) in absence external ac field
\cite{Sr}is 
seen in this model. SR is characterized by the observation of peak in
the particle mobility as function of system parameters such as $T$ and
$F$ in certain parameter space. The analysis of P\'{e}clet number in
this parameter space 
does not show any surprising features, so as to correlate
with SR.  This is due to the fact that SR occurs in the hight $T$ or
high $F$ regime, where barriers to motion are absent. To observe 
the peak in the P\'{e}clet number the existence of barrier seems to be 
essential. To clarify the relation between SR, NIS and P\'{e}clet number
one requires further detailed analysis.


\begin{thebibliography}{999}

\bibitem{Ris} H. Risken, {\em The Fokker Planck Equation}
    (Springer Verlag, Berlin, 1984).

\bibitem{Main} P. Reimann et al, Phys. Rev. E {\bf 65}, 031104
  (2002); P. Reimann et al, Phys. Rev. Lett {\bf 87}, 010602 (2001).

\bibitem{March} G. Constantini and F. Marchesoni,
  Europhys. Lett. {\bf 48}, 491 (1999).

\bibitem{Fnl} B. Lindner, M. Kostur and L. Schimansky-Geier,
  Fluct. Noise Lett. {\bf 1}, R25 (2001).

\bibitem{Geier} J. A. Freund and L. Schimansky-Geier, Phys. Rev. E
  {\bf 60}, 1304 (1999). 

\bibitem{Cr} A. Pikovsky and J. Kurths, Phys. Rev. Lett. {\bf 78}, 775
  (1997); B. Lindner and L. Schimansky-Geier, Phys. Rev. E {\bf 61},
  6103 (2000). 

\bibitem{Conf}
    Luc P. Faucheux and A. J. Libchaber,
    Phy. Rev. E {\bf49}, 5158 (1994).

\bibitem{Surf} H. Brenner,
    Chem. Eng. Sc. {\bf16}, 242 (1962).

\bibitem{Spc_mob} Rolf H. Luchsinger, Phys. Rev. E {\bf62}, 272 (2000).

\bibitem{Falco} C. M. Falco,
    Am. J. Phys. {\bf44}, 733 (1976). 


\bibitem{Sr} F. Marchesoni, Phys. Lett. A {\bf 231}, 61 (1997);
  D. Dan, M. C. Mahato and A. M. Jayannavar, Phys. Lett. A {\bf 258},
  217 (1999).
 

\bibitem{Dan} D. Dan, M. C. Mahato and A. M. Jayannavar, Phys. Rev. E
  {\bf 60}, 6421 (1999).


\bibitem{Danphys} D. Dan, M. C. Mahato and A. M. Jayannavar,
  Int. J. Mod. Phys. B {\bf 14}, 1585 (2000). 

\bibitem{Ratchet} Peter Reimann, Phys. Rep. {\bf 361}, 57 (2002);
  A. M. Jayannavar, cond-mat/0107079. 

\bibitem{Dan2} D. Dan, M. C. Mahato and A. M. Jayannavar, Phys. Rev. E
  {\bf 63}, 056307 (2001). 

\bibitem{Dan3} D. Dan and A. M. Jayannavar, Phys. Rev. E {\bf 65},
  037105 (2002). 


\bibitem{Langvn} M. C. Mahato and A. M. Jayannavar, Pramana-
 J. Phys {\bf 45}, 369 (1995); M. C. Mahato, T. P. Pareek and
 A. M. Jayannavar, Int. J. Mod. Phys B {\bf 10}, 3857 (1996).
 
\bibitem{Sancho} J. M. Sancho, M. San Miguel, and D. Duerr, J. Stat. Phys.
 {\bf 28}, 291 (1982).

\bibitem{Gard} C. W. Gardiner, {\em Handbook of Stochastic Methods}
  (Springer Verlag, Berlin, 1985).


\bibitem{Nis} R. N. Mantegna and B. Spagnolo, Phys. Rev. Lett {\bf 76},
563 (1996); M. C. Mahato and A. M. Jayannavar, Mod. Phys. Lett B {\bf 11},
815 (1997); M. C. Mahato and A. M. Jayannavar, Physica A {\bf 248},
138( 1998). 

\end{thebibliography}
\end{document}